\documentclass[vecphys]{svmult}

\usepackage{makeidx}         
\usepackage{graphicx}        
\usepackage{multicol}        
\usepackage{cite}            
\usepackage[bottom]{footmisc}
\usepackage{bbm}
\usepackage{bm}

\makeindex             

\newcommand{\ket}[1]{| #1 \rangle}
\newcommand{\bra}[1]{\langle #1 |}
\newcommand{\rb}[1]{\left( #1 \right)}
\newcommand{\ew}[1]{\langle #1 \rangle}
\newcommand{\beq}{\begin{eqnarray}}
\newcommand{\eeq}{\end{eqnarray}}

\newcommand{\eq}[1]{Eq.~(\ref{#1})}
\newcommand{\fig}[1]{Fig.~\ref{#1}}

\newcommand{\citer}[1]{{Ref.~\cite{#1}}}

\begin{document}

\title{Feedback control in quantum transport}
\author{Clive Emary}
\institute{
Department of Physics and Mathematics,
University of Hull,
Hull HU6 7RX,
United Kingdom
}

\maketitle

\begin{abstract}
Quantum transport is the study of the motion of electrons through nano-scale structures small enough that quantum effects are important.  In this contribution I review recent theoretical proposals to use the techniques of quantum feedback control to manipulate the properties of electron flows and states in quantum-transport devices.  Quantum control strategies can be grouped into two broad classes: measurement-based control and coherent control, and both are covered here.  I discuss how measurement-based techniques are capable of producing a range of effects, such as noise suppression, stabilisation of nonequillibrium quantum states and the realisation of a nano-electronic Maxwell's demon. I also describe recent results on coherent transport control and its relation to quantum networks.
\end{abstract}

\section{Introduction}
\label{SEC:intro}

Feedback control of quantum mechanical systems \index{quantum mechanics} is a rapidly emerging topic \cite{Gough2012,Zhang2015}, developed most fully in the field of quantum optics \cite{Wiseman2009}.   Only recently have these ideas been extended to quantum transport, a field which looks to understand and control the motion of electrons through structures on the nano-scale \cite{Nazarov2009}\index{quantum transport}. The aim of this contribution is to review these recent developments.

Broadly speaking, quantum feedback strategies may usefully be classified into two types:
\begin{itemize}
  \item {\em Measurement-based control}, where the quantum system is subject to measurements, the classical information from which forms the basis of the feedback loop;
  \item {\em Coherent control}, where the system, the controller and their interconnections are phase coherent such that the information flow in the feedback loop is of  quantum information \cite{Lloyd2000}.
\end{itemize}
Mirroring the situation in optics, most of the work to-date on feedback in quantum transport has been within the measurement-based paradigm.   In Sec.~\ref{SEC:mbc} here, we discuss a number of different measurement-based schemes and the physical results they can produce.  Initial studies of coherent control in quantum transport have recently been performed and these are discussed in Section~\ref{SEC:cc}.

\section{Measurement-based control \label{SEC:mbc}}

\begin{figure}[t]
  \begin{center}
    \includegraphics[width=0.8\columnwidth,clip=true]{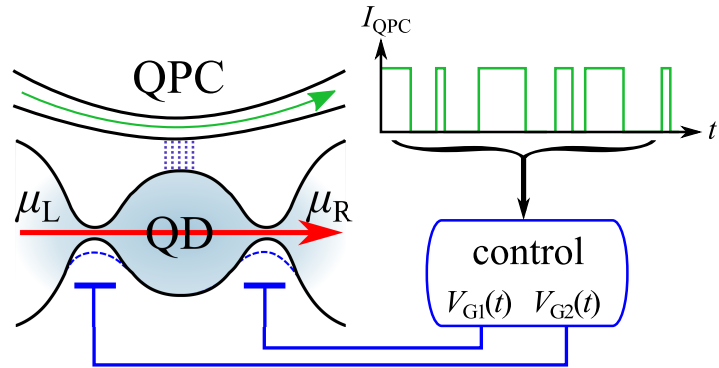}   
  \end{center}
  \caption{
    Schematic of a measurement-based feedback control scheme applied to transport through a quantum dot \index{quantum dot}(QD).
    The QD is connected to reservoirs (indicated by their chemical potentials $\mu_\mathrm{L}$ and $\mu_\mathrm{R}$) and the arrow indicates current flow.  The occupation status of the QD is detected with a quantum point contact (QPC), whose current gives rise to the time trace, top right.  This information is then processed by control circuitry that modulates the gate potentials $V_\mathrm{G1}(t)$ and $V_\mathrm{G2}(t)$ in response, and in doing so alters the tunnel rates of electrons through the QD.  In this way, a feedback loop is set up to control aspects of charge transfer through the dot.
    \label{FIG:MBcontrol}
  }
\end{figure}

The basic idea behind measurement-based control \index{measurement-based control} in quantum transport is sketched in \fig{FIG:MBcontrol}.
Generically, the transport system we are looking to control is a small quantum system in the Coulomb blockade regime, weakly coupled to leads across which a potential difference is applied.  In this regime, transport takes place via a series of discrete ``jumps'' in which electrons tunnel into or out of the system.

Our aim is to control some aspect of this process, be it the statistical properties of the current flow or the electronic states inside the device, through the establishment on a feedback loop based on the real-time detection of the electronic jumps using e.g. a  quantum point contact (QPC)\index{quantum point contact} \cite{Gustavsson2006,Fujisawa2006}.  The information gained from this electron counting is processed and used to e.g. manipulate the gate voltages that serve to define the various properties of the transport system such as the tunnel coupling between dot and reservoirs.  It should be noted that, although the system in question may be a quantum-mechanical one, the feedback loop here is an entirely classical affair.

\subsection{Counting statistics formalism}

The class of systems outlined above readily admits a description in terms of quantum master equations.  The full-counting statistics (FCS) \index{full-counting statistics} formalism \cite{Levitov1996} as applied to master equations \cite{Bagrets2003} provides a convenient way to calculate transport properties, as well as motivate, in a very physical way, various feedback schemes.

Let us consider a transport system described by a Markovian master equation\index{master equation}, 
\beq
  \dot \rho(t) = \mathcal{W}\rho(t)
  , 
\eeq
with $\rho(t)$ the reduced density matrix of the system at time $t$, and  $\mathcal{W}$ the Liouvillian\index{Liouvillian} superoperator of the system \cite{Brandes2005}. In a weak-coupling approach, $\mathcal{W}$ describes a simple rate equation with transitions between system eigenstates. 
Alternatively, in the infinite bias limit \cite{Gurvitz1996}, $\mathcal{W}$ defines a quantum master equation of Lindblad form with explicit unitary system dynamics and tunneling described in the local basis.

Irrespective of its precise form, the Liouvillian can be decomposed into terms that describe jump processes and those that do not.   Let us focus in on a single, particular jump process and decompose the Liouvillian as 
\beq
  \mathcal{W} = \mathcal{W}_0 + \mathcal{J}
  ,
\eeq
where $\mathcal{J}$ is the superoperator describing the jump process in question, and where $\mathcal{W}_0$ describes the remaining evolution without jumps of this kind.  
Defining $\rho^{(n)} (t)$ as the density matrix of the system conditioned on $n$ jumps of this type having occurred, the original master equation can be transformed into the number-resolved master equation \cite{Cook1981}
\beq
  \dot\rho^{(n)}(t) = \mathcal{W}_0\rho^{(n)}(t) +\mathcal{J} \rho^{(n-1)}(t) 
  \label{EQ:QMEnres}
  .
\eeq
Through definition of the  Fourier transform 
$
  \rho(\chi;t) \equiv 
  \sum_{n} 
  e^{i n\chi} 
  \rho^{(n)}(t)
$, we obtain the ``counting-field-resolved'' master equation
\beq
  \dot \rho(\chi;t) = \mathcal{W}(\chi)  \rho(\chi;t)
  ;\quad \quad
  \mathcal{W}(\chi) =\mathcal{W}_0 + e^{i\chi} \mathcal{J}
  \label{EQ:QMEchires}
  .
\eeq
This equation forms the basis of FCS calculations in master-equation approaches.
Generalisation to counting more than one type of transition is straightforward.

\subsection{Wiseman-Milburn control and the stabilisation of non-equillibrium pure states \label{SEC:WM}}

Perhaps the simplest quantum-control scheme, well understood in quantum optics, is that due to Wiseman and Milburn \index{Wiseman-Milburn control}\cite{Wiseman1994,Wiseman2009}.  In essence, this scheme monitors the quantum jumps of the system and, directly after each, applies a fixed control operation to the system, assumed to act instantaneously.  With such a control loop in place, counting and controlling a particular jump process, the dynamics of the system are still described by a master equation of the form \eq{EQ:QMEchires}, but with the original Liouvillian being replaced by its controlled counterpart
\beq
   \mathcal{W}(\chi) \to \mathcal{W}_C(\chi) 
   = \mathcal{W}_0  
   + e^{i\chi} \mathcal{C} \mathcal{J}
   \label{EQ:WMW}
   ,
\eeq
where $\mathcal{C}$ is the super-operator describing the control operation.  This operation is typically a unitary operation acting on the system, but could also include non-unitary elements (with possible changes to the counting field structure, e.g. \cite{Schaller2011}).

P\"{o}ltl {\em et al.} \cite{Poeltl2011} considered the application of Wiseman-Milburn control to transport models and demonstrated that it could be used to stabilise (in the sense to be defined below) a certain class of system state\index{state stabilisation}.
They considered a generic infinite-bias two-lead transport model with internal coherences, restricted to the zero or one charge sectors.  In this case the Liouvillian can be written as $\mathcal{W} = \mathcal{W}_0 +\mathcal{J}_L + \mathcal{J}_R$ with $\mathcal{J}_L$ describing electron tunneling in from the left and $\mathcal{J}_R$ describing tunneling out to the right.
The ``no-jump'' part of the Liouvillian can then be written in terms of a non-Hermitian Hamiltonian\index{non-Hermitian Hamiltonian}:
\beq
  {\cal W}_0 \rho = 
  -i
  \left\{
    \widetilde{H} \rho - \rho \widetilde{H}^\dag
  \right\}
  \label{free-ev}
  .
\eeq
This Hamiltonian has eigenstates
 $   \widetilde{H} \ket{\psi_j}= \varepsilon_j \ket{\psi_j} $ and 
$    \bra{\widetilde{\psi_j}} \widetilde{H}= \varepsilon_j \bra{\widetilde{\psi_j}}$, 
which, in general, are non-adjoint.
P\"{o}ltl {\em et al.} introduced a control operator $\mathcal{C}$ conditioned on the {\em incoming} jumps of the electrons and defined to rotate the post-jump state of the electron into one of the eigenstates $ \ket{\psi_j}$.
Since these states do not evolve under the action of $\mathcal{W}_0$, an electron in state $ \ket{\psi_j}$ will remain in it until it tunnels out.  The dynamics of the system with control can therefore described by a simple two-level model with effective Liouvillian (in the basis of populations of empty and $ \ket{\psi_j}$ states)
\beq 
  {\cal W}^{(j)}_C = 
  \rb{
  \begin{array}{cc}
    -\Gamma_{L} & \gamma_R^{(j)} \\
    \Gamma_L & - \gamma_R^{(j)}
  \end{array}
  }
  \label{EQ:Weff}
  ,
\eeq
where $\Gamma_{L} $ is the original rate of tunneling into the system and $\gamma_R^{(j)} = - 2 \,\,\mathrm{Im}(\varepsilon_j)$ is the new effective outgoing rate.   In the limit of high in-tunneling rate, $\Gamma_{L} \to \infty$, the system spends the majority of the time in state  $\ket{\psi_j}$ and the system is thus {\em stabilized} in this state.  
The state $\ket{\psi_j}$ is a {\em pure} state and thus very different from the stationary state of the system without control, which is typically mixed.  Furthermore, due to the non-equillibrium character of the effective Hamiltonian, these states are also distinct from the eigenstates of the original system Hamiltonian.
 
\citer{Poeltl2011} applied these ideas to a non-equillbrium charge qubit consisting of a double quantum dot \index{double quantum dot}with coherent interdot tunneling.  The stationary states of this model are mixed, to a greater or lesser degree, depending on the interdot tunnel coupling.  By using the above feedback scheme with the appropriate choice of unitary feedback operator, it was shown to be possible to stabilise states over the complete surface of the Bloch sphere.

This model was also used to illustrate the effects of the control scheme on the current flowing through the device.  
When exact stabilisation takes place and the system is governed by \eq{EQ:Weff}, the FCS of the system naturally reduces to that of a two-level system.  In the limit $\Gamma_L \to \infty$, these statistics become Poissonian, with all cumulants equal. This contrasts strongly with the FCS of the double quantum dot without control or with control parameters that do not lead to stabilisation. Thus, measurement of the output FCS can be used as part of a further (classical) feedback loop to isolate the stabilising control operation by minimising the distance between the system FCS distribution and that of a two-level system.

\subsection{Current-regulating feedback}

Historically, the first feedback control protocol to be proposed in quantum transport was that due to Brandes \cite{Brandes2010}, who considered a feedback loop which served to modify the various elements of \eq{EQ:QMEnres} such that they inherited a dependence on the number of jumps to have occurred:
\beq
  \dot\rho^{(n)}(t) = \mathcal{W}^{(n)}_0\rho^{(n)}(t) + \mathcal{J}^{(n)} \rho^{(n-1)}(t) 
  \label{EQ:QMEnresFB}
  .
\eeq
In particular, Brandes considered that the new elements in \eq{EQ:QMEnresFB} were the same as without feedback, but multiplied by analytic functions of the form
\beq
  f[q_n(t)];\quad q_n(t)\equiv I_{0} t - n;\quad f[0]=1
  \label{EQ:govern}
  .
\eeq
Here, the quantity $q_n(t)$ describes the deviation between the actual number of charges to have flowed through the system, $n$, and a reference, defined in terms of a reference current, $I_0$.  In the linear feedback case, $f(x)=1+gx$ with $g$ a small, dimensionless feedback parameter. Any rate multiplied by this function will increase if the actual charge transferred lags behind the reference, and decrease if it is in excess\index{current regulation}.

\begin{figure}[t]
  \begin{center}
    \includegraphics[width=0.75\columnwidth,clip=true]{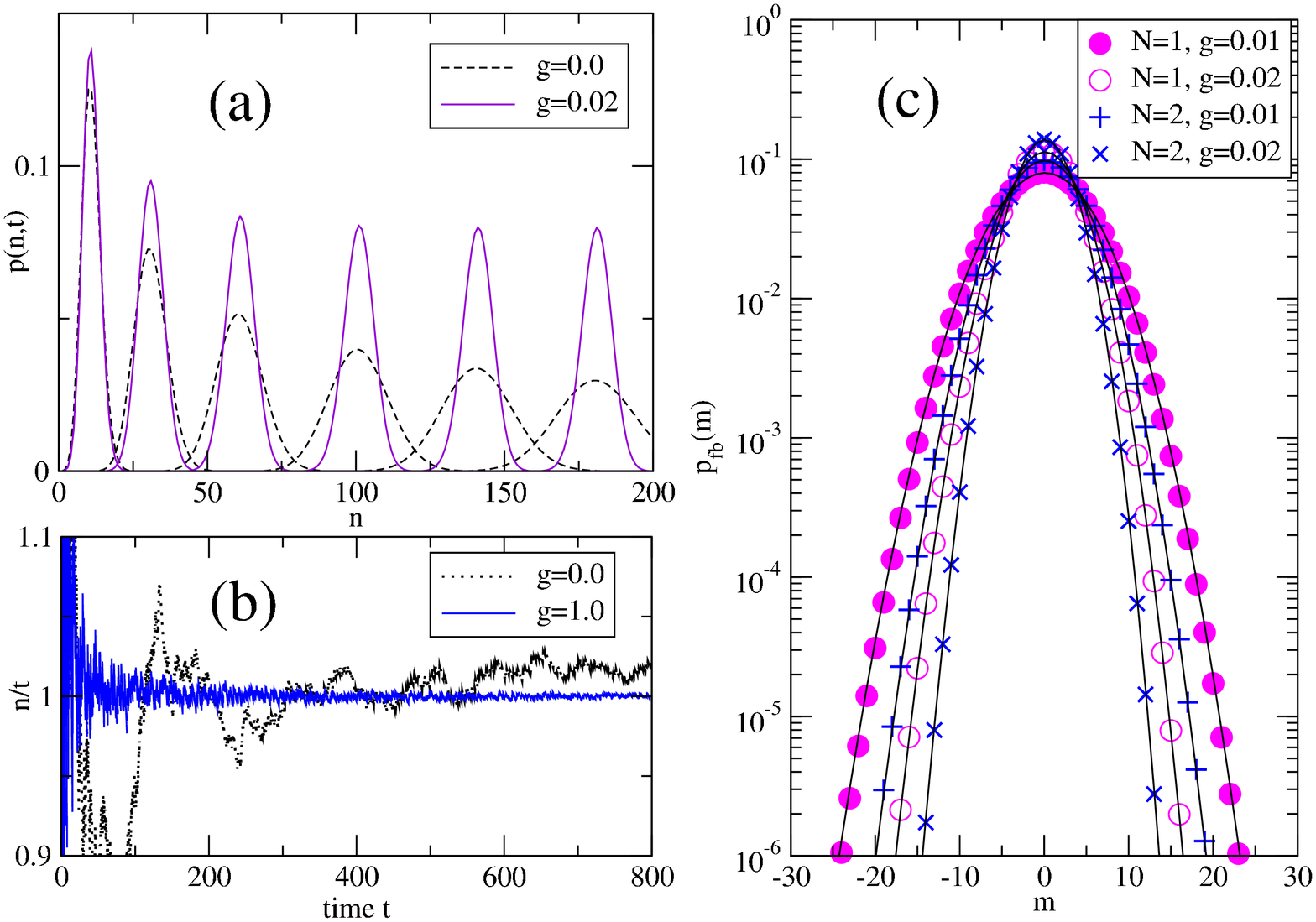}   
  \end{center}
  \caption{
    The effect of current-governing feedback.  Here is shown the distribution of the number of electrons $n$ transferred through a tunnel junction
    at times $t = 30, 60, 100, 140, 180$  ($\Gamma=1$).  Shown are results for the case without ($g=0$) and with linear ($g=0.02$) feedback.  Inclusion of the current-governing feedback leads to a freezing of the distribution. From \citer{Brandes2010}.
    \label{FIG:Brandes}
  }
\end{figure}

The results obtained with this current-regulating feedback are exemplified by the simple model of a unidirectional tunnel junction where $\mathcal{J}^{(n)} = -\mathcal{W}^{(n)}_0 = \Gamma\left\{1 + g \rb{I_0t - n}\right\}$ is scalar.  Without control, this is a Poisson process and all cumulants of the transfered-charge distribution are equal: $C_k(t) = \Gamma t$. In particular, the width of the distribution grows linearly with time: $C_2(t) \equiv \ew{n^2}(t) - \ew{n}^2(t)= \Gamma t$.  With control in place, and $I_0$ set as the mean current without control, the first cumulant is unaltered.  However, the second cumulant becomes
\beq
 C_2(t) = \ew{n^2}(t) - \ew{n}^2(t) = \frac{1}{2g} \rb{1-e^{-2g\Gamma t}}
 .
\eeq
Thus, the width of the FCS distribution no longer grows in time, but rather tends towards a fixed value.  \fig{FIG:Brandes} shows the FCS distribution with and without control and illustrates that feedback of this type leads to a freezing of the shape of the distribution. 
In the long-time limit, then, the relative fluctuation in the electron number,  becomes vanishing small since $C_2(t)/C_1(t) \sim (2 g \Gamma t)^{-1}$.
This effect was also shown to hold for higher-dimensional models, in particular the single-electron transistor.

One potential application of this effect might be in the control of single-electron current sources \cite{Pekola2013}, where reduction of the fluctuations in electron current is essential for the realisation of a useful quantum-mechanical definition of the ampere \cite{Giblin2012}.  In this context, Fricke et al. have demonstrated the current locking of two electron pumps through a feedback mechanism based on the charging an mesoscopic island between them \cite{Fricke2011}\index{electron pump}.

Whilst the number-dependence of \eq{EQ:QMEnresFB} was originally considered to be the result of some external classical feedback circuit, 
subsequent work has shown that this kind of dependence can also arise from microscopic considerations \cite{Schubotz2011}.  Recently, Brandes has shown how the feedback coupling of \eq{EQ:govern} can arise autonomously in a series of interacting transport channels \cite{Brandes2015}.

\subsection{Piecewise-constant feedback and Maxwell's demon}

The idea of piecewise-constant feedback \index{piecewise-constant feedback}is best illustrated by direct consideration of the Maxwell's demon \index{Maxwell's demon} proposal of \citer{Schaller2011}.  The set-up is exactly as in \fig{FIG:MBcontrol} with the quantum dot restricted to just two states (`empty' and `full') and with the two reservoirs at finite bias and temperature.

The population of the QD is monitored in real-time and with piecewise-constant feedback we apply one configuration of top-gate voltages when the QD is occupied, and a different set when it is empty.  The electrons thus experience a different Liouvillian depending on whether the QD is empty of full. As shown in \citer{Schaller2011, Schaller2014}, however, it is possible to describe the evolution of the system in terms of a single Liouvillian which, in the current case, written in the basis populations (empty, full), reads:
\beq
  \mathcal{W}_{\rm fb}^I(\chi_L,\chi_R) &=& 
  \mathcal{W}_E (\chi_L,\chi_R)
  \left(\begin{array}{cc}
  1 & 0\\
  0 & 0
  \end{array}\right)
  +\mathcal{W}_F(\chi_L,\chi_R)
  \left(\begin{array}{cc}
  0 & 0\\
  0 & 1
  \end{array}\right)\,.
  \label{EQ:EfbI}
\eeq
Here $\mathcal{W}_E$ is the Liouvillian when system in empty, and $\mathcal{W}_F(\chi_L,\chi_R)$ the Liouvillian when full.  Both left ($\chi_L$) and right ($\chi_R$) counting fields are required.

In \citer{Schaller2011} it was assumed that only the dot-lead tunneling rates (and not, for example, the position of the energy level in the dot) are changed by the feedback loop.  It can easily be seen how this arrangement might lead to a Maxwell's demon if we imagine that when the dot is empty, we completely close off tunneling to the right; and when the dot is full, we reverse the situation, and close tunneling to the left. In this situation, irrespective of bias or temperature, electrons will be preferentially transported from left to right through the QD. 
Even with less extreme modulation of the barriers, this scheme was shown to still  drive a current against an applied bias, and thus extract work. In the classical limit, changing the barriers in the above fashion performs no work on the system, and thus, the current flow arises from the information gain of the feedback loop.  This then is equivalent to Maxwell's demon.
Two further feedback schemes, both based on Wiseman-Milburn style instantaneous control pulses, were also considered and shown to also give rise to the demon effect.

Subsequently Esposito and Schaller \cite{Esposito2012} have formalised the notion of ``Maxwell demon feedbacks'' and studied their thermodynamics.  A physical, autonomous implementation of these ideas was discussed in \citer{Strasberg2013}, where the demon was realised by a second quantum dot connected to an independent electron reservoir.
We note also that a similar proposal involving a three-junction electron pump has also been proposed \cite{Averin2011}.

\subsection{Feedback control with delay}

The preceding schemes have all assumed that the control operation is effected on the system immediately after the detection of a jump.  In reality, however, there will always be a delay, of a time $\tau$ say, between detection and actuation \index{delay}.
Wiseman considered the effects of delay on the class of feedback schemes outlined in Sec.~\ref{SEC:WM} and gave modifications to \eq{EQ:WMW} correct to first order in the delay time \cite{Wiseman1994}.   In \citer{Emary2013b}, I showed that Wiseman's result actually holds for arbitrary delay time, providing one makes an additional ``control-skipping''\index{control-skipping assumption} assumption, which means that if a jump occurs within the delay-time of an earlier jump, then the control operation for the first jump is discarded.
With this assumption, it is straightforward to show that the delay-controlled system still obeys a master equation, but now a nonMarkovian one\index{nonMarkovian master equation}.  The appropriate replacement for the kernel reads
\beq
   \mathcal{W}(\chi) \to
   \mathcal{W}_{DC}(\chi,z) 
   &=& 
  \mathcal{W}_0 
  +
    \mathcal{D}(\chi,z) 
    \mathcal{J} e^{i\chi}
    \label{EQ:WDCtwoJ}
    ,
\eeq
with
\beq
  \mathcal{D}(\chi,z) 
  &=&  
  \mathbbm{1} + \left[\mathcal{C}-\mathbbm{1}\right]e^{(\mathcal{W}_0-z)\tau} 
  \label{EQ:DztwoJ}
  ,
\eeq
the delayed control operation.  In these expressions, $z$ is the variable conjugate to time in the Laplace transform.  In the time domain, we obtain the delayed nonMarkovian master equation
\beq
  \dot\rho(t) =
    \mathcal{W}\rho(t) 
    +
    (\mathcal{C}-1) e^{\mathcal{W}_0\tau}\mathcal{J}
     \rho(t-\tau)  \theta(t-\tau)
  \label{NMMEtime}
    ,
\eeq
in which the time evolution of the density matrix $\rho(t)$ depends not only on the state of the system at time $t$ but also at previous time $t-\tau$.

\citer{Emary2013b} considered the effect of delay on the state-stabilisation protocol of \citer{Poeltl2011} and the Maxwell's demon of \citer{Schaller2011}.  The influence of delay on the current governor was also discussed in \cite{Brandes2010}.  In all these cases, the effects of delay are deleterious, but some effect of the control loop persists in the presence of delay.  The influence of delay on the thermodynamics of Wiseman-Milburn feedback was studied in \citer{Strasberg2013a}.

\section{Coherent control \label{SEC:cc}}

Coherent feedback seeks to control a quantum system without the additional disturbance produced by the measurement step in measurement-based control.  Various forms of coherent control have been discussed in the literature, e.g. Refs.~\cite{Lloyd2000,Mabuchi2008,Zhang2011}. However, the only type currently proposed for quantum transport \cite{Emary2014a,Gough2014} is the {\em quantum feedback network} \cite{Gough2008,Gough2009,Gough2009a,James2008,Nurdin2009,Zhang2011,Zhang2012}, and this is the work we describe here\index{coherent control}.

\subsection{Quantum feedback networks}

In contrast to the measurement-based case, the quantum-feedback-network approach of \citer{Emary2014a} assumes that the system is strongly coupled to the leads, that the motion of the electrons through system and controller is phase coherent and that electron-electron interactions can be neglected.  In this limit, transport can be described by Landauer-B\"uttiker theory \cite{Blanter2000}\index{quantum feedback network}\index{Landauer-B\"uttiker theory}.

\begin{figure}[t]
  \begin{center}
     \includegraphics[width=0.8\columnwidth,clip=true]{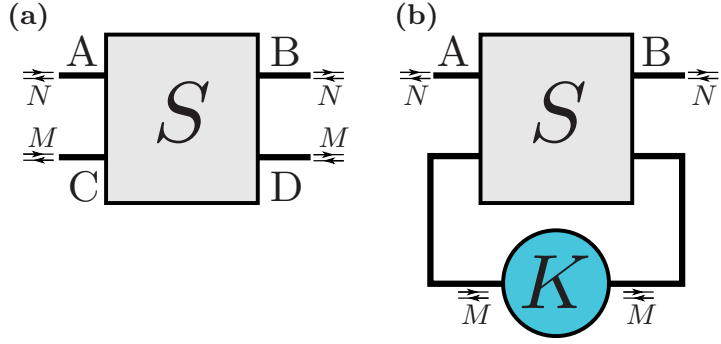}
  \end{center}
  \caption{
    \textbf{(a)}:
    At the centre of the quantum feedback network discussed in \citer{Emary2014a} is a four-terminal device with scattering matrix $S$.  Four leads are labeled A through D: A and B possess $N$ bidirectional channels and leads C and D possess $M$.
    \textbf{(b)}:
    The feedback loop is realised by connecting leads C and D together  via a controller with scattering matrix, $K$.
    Figure taken from \citer{Emary2014a}.
    \label{FIG:SFB}
  }
\end{figure}

\citer{Emary2014a} considered that the system to controlled was a four-terminal device (see \fig{FIG:SFB}), whose scattering matrix \index{scattering matrix}could be written in block form as
\beq
  S = 
  \rb{
    \begin{array}{cc}
      S_\mathrm{I} & S_\mathrm{II} \\
      S_\mathrm{III} & S_\mathrm{IV}
    \end{array}
  }
  \label{EQ:Sfull}
  .
\eeq
Here, for example, block $ S_\mathrm{I}$ describes all scattering processes between leads A and B, $ S_\mathrm{II}$ describes all scattering processes starting in leads C and D and ending in leads A and B, and so on.  All leads support states travelling in both directions.
The feedback network is then formed by connecting leads C and D together through a controller with scattering matrix $K$. By considering all possible paths between leads A and B, the joint scattering matrix for the system-controller network is found to be:
\beq
  S_\mathrm{fb}
  =
  S_\mathrm{I}
  +
  S_\mathrm{II}
  \frac{1}{\mathbbm{1}- K S_\mathrm{IV}}
  K S_\mathrm{III}
  \label{EQ:SFB1}
  .
\eeq

One key result stemming from this is that if the control scattering matrix, $K$, has the same dimension as the output matrix, $S_\mathrm{fb}$, we can rearrange \eq{EQ:SFB1} to give\index{ideal control}
\beq
  K = 
  \frac{1}{S_\mathrm{IV} + S_\mathrm{III} \rb{S_\mathrm{fb}-S_\mathrm{I}}^{-1}S_\mathrm{II}}
  \label{EQ:Kideal}
  .
\eeq
Thus, given an arbitrary original system matrix, $S$, we can obtain any desired output $S_\mathrm{fb}$ by choosing the control operator as in \eq{EQ:Kideal}.  This was dubbed ``ideal control'' in \citer{Emary2014a}.

\subsection{Conductance optimisation}

\begin{figure}[t]
  \begin{center}
    \includegraphics[width=\columnwidth,clip=true]{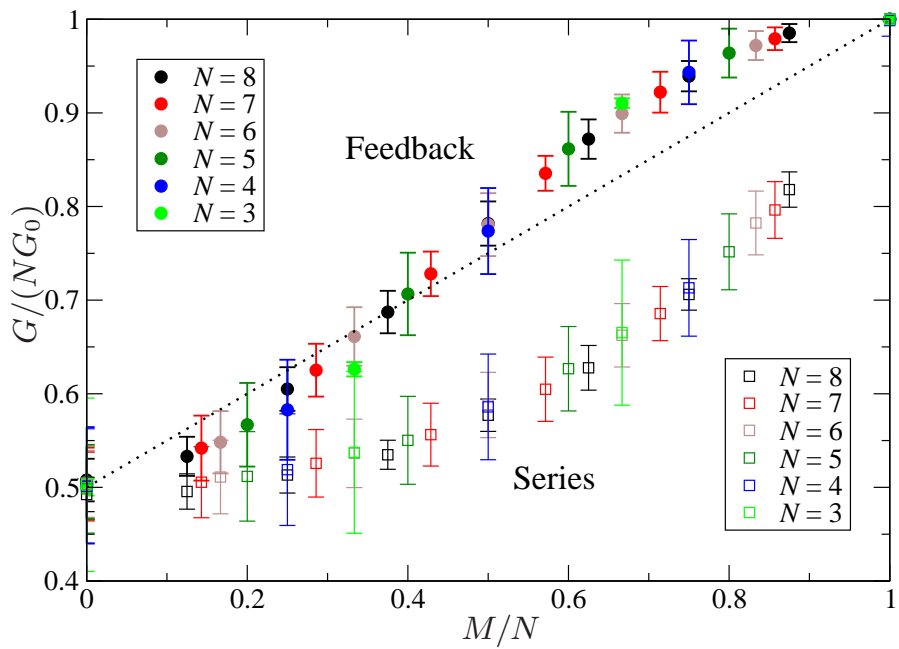}   
  \end{center}
  \caption{
    Optimised conductance $G$ of an ensemble of chaotic quantum dots under coherent control plotted as a function of the ratio of control to output dimension, $M/N$.  The conductance is given in units of the ballistic conductance, $NG_0$, with $G_0= 2e^2 / h$.  Results are shown for both feedback (solid circles) and series (open squares) configurations.  Increasing the controller dimension increases the conductance gain towards the ideal-control limit of $G =NG_0$ at $M=N$. The feedback geometry outperforms its series counterpart for all $0<M<N$. From \citer{Emary2014a}.
    \label{FIG:QDFBconductance}
  }
\end{figure}

As an example of the use of the feedback network described by \eq{EQ:SFB1}, \citer{Emary2014a} studied the optimisation of the conductance of chaotic quantum dots\index{conductance optimisation}\index{quantum dot}.  The dots were modeled with $4N\times4N$ scattering matrices taken from random matrix theory \cite{Beenakker1997}.  The number of active control channels in the controller was set as $1 \le M \le N$ and the elements of $K$ chosen to maximise the conductance of the system-controller network.
Results for the feedback network are shown in \fig{FIG:QDFBconductance}, and compared with those for a second network where system and controller are placed in series. Without control ($M=0$), the conductance is given by the random-matrix-theory result $G/(N G_0) = 1/2$, with $G_0$ the conductance quantum. 
When $M=N$, ideal control is possible for both series and feedback setups and the ballistic conductance $G/(N G_0) = 1$ is obtained.
For $0 < M < N$ there is a monotonic increase in the conductance for both series and feedback geometries.  However, it is the feedback loop that offers the greater degree of conductance increase. The calculations of \citer{Emary2014a}  also indicate that the feedback-loop geometry is also more robust under the influence of decoherence\index{decoherence}.

\section{Conclusion}
\label{SEC:conclusion}

In the majority of the proposals reviewed here, the target of the control has been the current flowing through the device.  We have seen ways in which either the magnitude of the current, as in the Maxwell-demon and coherent-control proposals, or its statistical properties, as in the governor, can be modified by feedback. The exception to this was the proposal in Sec.~\ref{SEC:WM}, where the target of the control was the nonequillibrium states of the electrons inside the system itself.  The manipulation of these states, however, also had a knock-on effect for current, which proved useful in diagnosing the effectiveness of the control procedure.

Of these proposals, the current-governor and Maxwell demon are the most hopeful candidates for experimental realisation in the foreseeable future.
Indeed, a Maxwell's demon, similar in many respects to the one described here, has recently been realised in the single-electron box \cite{Koski2014}.
Such schemes are practicable because, although they take place in quantum-confined nanostructures, they do not rely on quantum coherence for their operation.  The time-scales involved can therefore be relatively slow: in the FCS experiments of \citer{Gustavsson2006}, for example, the QD was very weakly coupled to the reservoirs so that the typical time between tunnel events was of the order of a millisecond.  Given this sort of timescale, it should be possible to build a control circuit fast enough to enact the required operations with a speed approximating the instantaneous ideal. 
By way of contrast, the pure-state stabilisation proposal of \citer{Poeltl2011} only makes sense  if the control loop operates on a timescale faster than the coherence time of the system being controlled.  For charge coherences, this time is $\sim 1$ns \cite{Fujisawa2006a}, rending the construction of the feedback loop a considerable challenge.  
Progress could perhaps be made by controlling spin, rather than charge, degrees of freedom, since spin coherence times in QDs are far longer\index{coherence times}.
We note that all such questions of external-circuits timescale are side-stepped by coherent-control protocols such as that described in Sec.~\ref{SEC:cc}.  Here, however, the challenge is to go beyond abstract analysis and find appropriate physical systems to act as useful controllers.

\bibliographystyle{spphys.bst}


\end{document}